\documentclass[11pt]{article}

\usepackage[preprint]{acl}

\usepackage{times}
\usepackage{latexsym}
\usepackage[T1]{fontenc}
\usepackage[utf8]{inputenc}
\usepackage{microtype}
\usepackage{inconsolata}

\usepackage{enumitem}
\usepackage{multirow}
\usepackage{colortbl}
\definecolor{mygray}{gray}{0.95}
\usepackage{float}
\definecolor{bgcolor}{rgb}{0.9, 0.95, 1.0}
\usepackage{xcolor}
\usepackage{amsmath}
\usepackage{amssymb}
\usepackage{mathtools}
\usepackage{amsthm}
\usepackage{placeins}
\usepackage[most]{tcolorbox}
\tcbset{colback=white, colframe=black!40, fonttitle=\bfseries, breakable}
\usepackage{graphicx}
\usepackage{subcaption}
\usepackage{booktabs}
\usepackage{algorithm}
\usepackage{algorithmic}
\usepackage{hyperref}
\usepackage[capitalize,noabbrev]{cleveref}

\usepackage{makecell}
\usepackage{xcolor}
\usepackage{colortbl}

\newcommand{\RETURN}{\STATE \textbf{return}~}  

\usepackage{xcolor}

\usepackage{cuted}

\theoremstyle{plain}

\theoremstyle{definition}

\theoremstyle{remark}

\title{EviProp: Seeded Relevance Diffusion on Chunk–Page Graphs for Long Multimodal Document Retrieval}

\author{
\textbf{Hongwei Zhang}\textsuperscript{1,2},
\textbf{Xiaoman Wang}\textsuperscript{1},
\textbf{Zehui Ling}\textsuperscript{3},
\textbf{Ruicheng Zhu}\textsuperscript{4},
\textbf{Yue Zhang}\textsuperscript{2,5},\\
\textbf{Pinlong Cai}\textsuperscript{2},
\textbf{Fuke Shen}\textsuperscript{1},
\textbf{Botian Shi}\textsuperscript{2},
\textbf{Tongquan Wei}\textsuperscript{1}\thanks{Corresponding authors.},
\textbf{Guohang Yan}\textsuperscript{2}\footnotemark[1]
\\
\textsuperscript{1}East China Normal University,
\textsuperscript{2}Shanghai Artificial Intelligence Laboratory, \\
\textsuperscript{3}Fudan University,
\textsuperscript{4}Shanghai Jiao Tong University,
\textsuperscript{5}University of Shanghai for Science and Technology
}

\begin{document}
\maketitle

\begin{abstract}
Retrieving evidence pages from visually rich long documents is a key challenge in document question answering.
Existing page-level visual retrievers operate under an independent matching paradigm: each page is scored in isolation 
based on query--page similarity.
This paradigm can under-rank evidence pages whose signals are localized in fine-grained chunks or depend on document-internal associations.
We propose \textbf{EviProp}, a retrieval method that recovers such pages via seeded relevance diffusion.
EviProp models each document as a multimodal Chunk--Page graph with hierarchical, sequential, and similarity links.
Given a query, it combines dense visual page priors with sparse chunk seeds, then runs Personalized PageRank to diffuse relevance over the graph.
Experiments on MMLongBench-Doc and LongDocURL show consistent gains in evidence-page retrieval over independent visual 
retrieval and text--visual fusion baselines. Downstream QA results further show that improved retrieval translates into better answer accuracy, with negligible online retrieval overhead. Our code is released at \url{https://github.com/Flyecnu/EviProp}.
\end{abstract}

\section{Introduction}
\label{sec:intro}

\begin{figure}[t]
  \centering
  \includegraphics[width=0.9\linewidth]{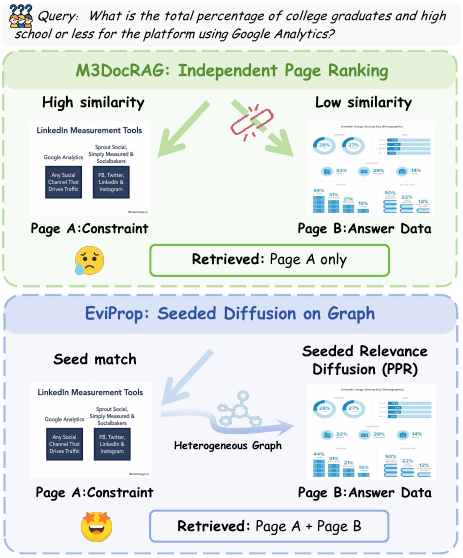}
  \caption{
  \textbf{Motivating example on MMLongBench-Doc.}
  Independent page ranking retrieves only the constraint page and misses the evidence page.
  \textbf{EviProp} recovers the evidence page by diffusing relevance over a Chunk--Page graph initialized with page priors and chunk seeds.}
  \label{fig:motivation}
\end{figure}

Document understanding is foundational to real-world applications spanning financial analysis and scientific literature review~\cite{V-doc,suri2025visdom,zhang2025ocr}.
The field has shifted from OCR-based text pipelines toward vision-native processing with Large Vision-Language Models 
(LVLMs)~\cite{mmlongbench,longdocurl}, which interpret document images directly and preserve charts, tables, and complex layouts.
However, directly applying LVLMs to all pages of lengthy real-world documents remains challenging due to context-window 
limitations and high inference costs. Consequently, retrieving the most relevant \textit{evidence pages} first has become the standard practice in LVLM-based document QA~\cite{m3docrag,mdocagent}, providing concise visual context for downstream reasoning.

Current retrievers such as ColPali~\cite{colpali} score each page independently by global query--page visual similarity.
While effective for coarse retrieval, this independent matching paradigm remains fundamentally limited for fine-grained evidence discovery in long documents, for two reasons.
First, \textbf{localized evidence is diluted}: answer-supporting content such as a sparse numerical statistic, a table cell, or a chart region often occupies only a small fraction of a page.
As a result, the page's global visual similarity to the query remains weak even when it contains the decisive evidence, causing it to be systematically under-ranked.
Second, \textbf{document-internal structure is ignored}: page-wise retrievers treat each page as an isolated unit, discarding structural and semantic associations among pages and fine-grained content units.
Evidence supported by neighboring pages, semantically related regions, or cross-page document structures therefore cannot be effectively recovered during retrieval.
Together, these limitations cause evidence pages whose relevance is latent or structurally distributed to be persistently under-ranked by independent page matching.

As illustrated in Figure~\ref{fig:motivation}, a representative failure case from MMLongBench-Doc~\cite{mmlongbench} highlights this limitation.
The retriever successfully retrieves the page explicitly mentioning ``Google Analytics'' (Page~A), but misses the actual evidence page containing the required education statistics (Page~B) because the evidence page does not strongly match the query globally.
Once the evidence page is excluded during retrieval, the downstream LVLM never observes the necessary information and therefore fails to answer correctly.
More broadly, this issue is not limited to multi-page reasoning: even for single-page questions, independent page ranking can still under-rank the correct page when decisive evidence is sparse or localized.

To address this limitation, we propose \textbf{EviProp}, a retrieval method that improves evidence-page discovery via \textbf{seeded relevance diffusion}. Starting from a visual retriever, EviProp parses each document into fine-grained textual chunks and visual regions, constructing a \textbf{multimodal Chunk--Page heterogeneous graph}. The graph explicitly encodes (1) chunk--page membership, (2) page-level sequential continuity, and (3) semantic and visual similarity links across chunks and pages. Given a query, EviProp initializes a restart distribution by combining \textbf{dense visual priors} (from global page matching) with \textbf{sparse chunk seeds} (from precise local matching). We then perform Personalized PageRank (PPR)~\cite{ppr} to diffuse relevance over the graph. This process allows local chunk evidence, page-level visual priors, and document-internal associations to jointly influence page relevance, thereby promoting evidence-bearing pages that are under-ranked by independent page matching.

Our contributions are summarized as follows:
\begin{itemize}[leftmargin=1.2em, itemsep=0.2em, topsep=0.2em]
    \item We propose EviProp, a retrieval method that reformulates evidence-page retrieval as seeded relevance diffusion over a multimodal Chunk--Page graph, addressing the limitations of independent page-level matching.
    
    \item EviProp introduces a sparse-dense seeding strategy that combines dense page-level visual priors with sparse chunk seeds, allowing localized evidence signals to propagate through hierarchical, sequential, and similarity-based document relations.

    \item Experiments on MMLongBench-Doc and LongDocURL show consistent gains in evidence-page retrieval over visual and hybrid retrieval baselines. Downstream QA results further show that improved retrieval translates into better answer accuracy with negligible online overhead.
\end{itemize}

\section{Related Work}
\label{sec:related_work}

\subsection{Multimodal Document Retrieval}
DocQA has evolved from single-page or short-document analysis~\cite{docvqa,tito2023hierarchical,slidevqa} to lengthy multimodal documents~\cite{mmlongbench,longdocurl}. Vision-native LVLM methods directly process page images~\cite{dse,vdocrag}, but applying them to long documents requires evidence-page retrieval to reduce input burden~\cite{nldir}. Methods such as ColPali~\cite{colpali}, VisRAG~\cite{visrag} and M3DocRAG~\cite{m3docrag} advance vision-native page retrieval, while MDocAgent~\cite{mdocagent} combines text and image agents for collaborative QA. However, these methods commonly rely on independent page-level matching, scoring each page in isolation. This can under-rank evidence pages whose signals are localized in fine-grained chunks or depend on document-internal associations. EviProp addresses this limitation by explicitly propagating relevance across document structures rather than scoring pages in isolation.

\subsection{Graph-based Retrieval and Relevance Propagation}

Graph-based retrieval has been explored for multi-hop reasoning in RAG systems~\cite{graphrag,lightrag,hirag}. For multimodal documents, RAG-Anything~\cite{raganything} and related methods~\cite{aligning,mkg,ladrag} textualize visual elements to construct graph structures, relying on expensive LLM-based relation extraction. MoLoRAG~\cite{molorag} constructs a page similarity graph and uses a VLM to score visited pages. Nevertheless, page relevance is still estimated independently during traversal, without explicit relevance propagation across fine-grained document structures. 
In the text domain, HippoRAG~\cite{hipporag,hipporag2} and LinearRAG~\cite{linearrag} demonstrate that Personalized PageRank~\cite{ppr} supports efficient associative retrieval. However, these methods are restricted to pure text corpora and cannot process visually-rich documents or leverage multimodal page priors. EviProp instead performs seeded relevance diffusion over a multimodal Chunk--Page graph, combining dense visual page priors with sparse chunk seeds to recover evidence pages without invoking a VLM at retrieval time.

\section{Method}
\label{sec:method}

\subsection{Problem Definition}
\label{sec:setup}

Standard visual retrievers score each page in isolation, making it difficult to recover evidence pages whose signals are localized in fine-grained chunks or depend on document-internal associations. EviProp focuses on evidence-page retrieval for visually rich long documents.
Let $D$ be a document consisting of an ordered sequence of $M$ pages, $\mathcal{P}=\{p_1, \dots, p_M\}$.
Each page $p_i$ is a multimodal unit containing text, layout structures, and visual figures.
Given a user query $Q$, our goal is to rank pages by their likelihood of containing answer-supporting evidence and return the top-$K$ pages $\mathcal{P}_r \subset \mathcal{P}$ ($K \ll M$).
The retrieved pages are then used as visual context for a downstream Large Vision-Language Model (LVLM), while our method focuses on improving the retrieval stage.
This setting includes cases where evidence is sparse on a single page or supported by document-internal associations.

\begin{figure*}[t] 
  \centering
  \includegraphics[width=0.98\textwidth]{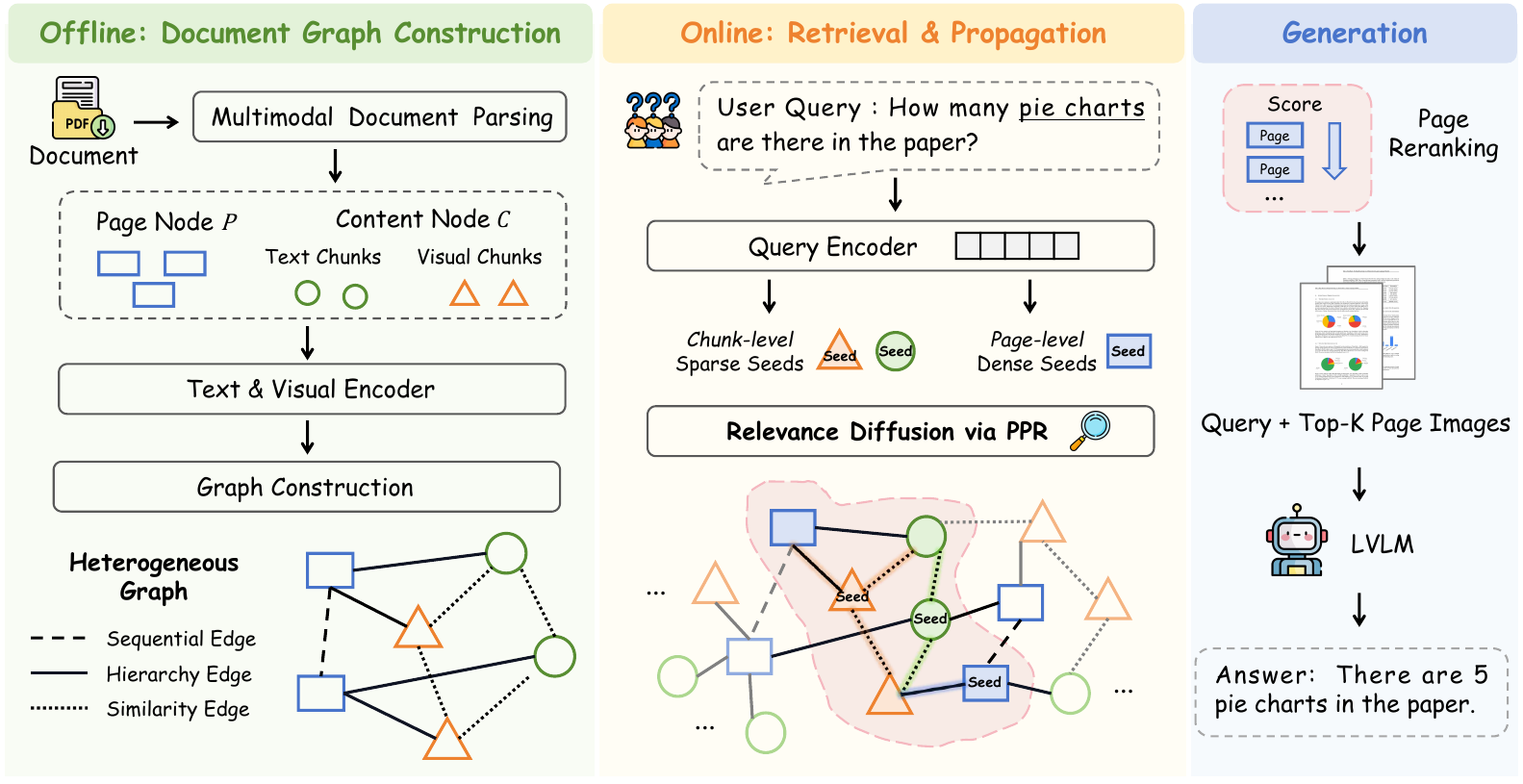}
  \caption{
  \textbf{Illustration of \textbf{EviProp}.}
  \textbf{EviProp} constructs a Chunk--Page graph offline and performs seeded relevance diffusion online to retrieve evidence pages.
  The downstream LVLM uses the retrieved pages as visual context for answer generation, but the generation module is not modified by \textbf{EviProp}.
  }
  \label{fig:overview}
\end{figure*}

\subsection{Overview of EviProp}
As illustrated in Figure~\ref{fig:overview}, EviProp operates in four stages:
(1) \textbf{Chunk--Page Graph Construction}, building a multimodal Chunk--Page graph offline;
(2) \textbf{Sparse-Dense Evidence Seeding}, initializing node relevance from dense page priors and sparse chunk seeds;
(3) \textbf{Relevance Diffusion via Personalized PageRank}, propagating relevance via Personalized PageRank;
and (4) \textbf{Final Evidence-Page Scoring}, combining diffused posteriors with visual priors for page ranking.

\subsection{Chunk--Page Graph Construction}

\label{sec:graph_construction}

Given a visually rich long document
$\mathcal{D}=\{p_1,\dots,p_M\}$,
EviProp constructs a multimodal heterogeneous graph
$\mathcal{G}=(\mathcal{V},\mathcal{E})$
to support relevance propagation across local evidence, sequential page context, and non-local document associations.

\paragraph{Node Construction.}
We first perform multimodal document parsing using MinerU~\cite{mineru} to extract fine-grained content units from each page.
The node set
$\mathcal{V}=\mathcal{P}\cup\mathcal{C}$
consists of:
\begin{itemize}[leftmargin=*, nosep]
    \item \textbf{Page Nodes ($\mathcal{P}$):}
    Each page $p_i$ is represented as a global visual context node.

    \item \textbf{Text Chunk Nodes ($\mathcal{C}_{\mathrm{text}}$):}
    Derived from OCR text lines or paragraphs.

    \item \textbf{Visual Chunk Nodes ($\mathcal{C}_{\mathrm{vis}}$):}
    Corresponding to cropped table, chart, or figure regions.
    Each visual chunk is further augmented with an offline-generated caption to bridge the modality gap between visual regions and textual queries.
\end{itemize}

The complete content node set is:
\begin{equation}
\mathcal{C}
=
\mathcal{C}_{\mathrm{text}}
\cup
\mathcal{C}_{\mathrm{vis}}.
\end{equation}

\paragraph{Edge Construction.}
We define a weighted edge function
\(
w(u,v)
\)
to measure the transition strength between two connected nodes
\(u,v \in \mathcal{V}\). Based on document structure and semantic associations, we construct three types of edges.

\begin{itemize}[leftmargin=*,nosep]

\item \textbf{Hierarchical Membership Edges.} Each content node $c_j$ is connected to its parent page $p_i$ via bidirectional edges:
\begin{equation}
w(c_j,p_i)=w(p_i,c_j)=w_{\mathrm{inc}}.
\end{equation}
where $w_{\mathrm{inc}}$ is a constant membership weight.
These edges allow localized chunk evidence to directly influence page relevance. 

\item \textbf{Sequential Page Edges.} To preserve document continuity, adjacent pages are connected by bidirectional edges:
\begin{equation}
w(p_i,p_{i+1})
=
w(p_{i+1},p_i)
=
w_{\mathrm{seq}}.
\end{equation}
where $w_{\mathrm{seq}}$ controls the propagation strength between neighboring pages.
This enables relevance propagation across neighboring document contexts.

\item \textbf{Similarity Edges.} To capture non-local associations, we introduce similarity-based edges across pages and chunks.

For page nodes, visual similarity is computed by mean-pooling the multi-vector representation $\mathbf{V}_{p_i}$ into a single vector and computing cosine similarity:
\begin{equation}
w(p_i,p_j) = w(p_j,p_i) 
= \max\!\left(0,\, \bar{\mathbf{v}}_{p_i}^\top \bar{\mathbf{v}}_{p_j}\right),
\end{equation}
where negative similarities are clipped to zero to suppress spurious associations.

For chunk nodes, to suppress noise from weak semantic overlaps and prioritize strong associations, we assign the edge weight using cubic scaling:
\begin{equation}
w(c_i,c_j) = w(c_j,c_i) 
= \cos(\mathbf{t}_{c_i},\mathbf{t}_{c_j})^3.
\end{equation}
Edges whose similarity falls below a threshold $\tau_{\mathrm{txt}}$ are discarded.\footnote{Detailed edge weighting schemes and hyperparameter settings are provided in Appendix~\ref{app:eviprop_settings} and~\ref{app:graph_weights}.}

\end{itemize}

\paragraph{Transition Matrix Construction.}

All edge weights are assembled into the weighted adjacency matrix
$\mathbf{W}$,
where
$W_{ij}=w(v_i,v_j)$.
We then row-normalize $\mathbf{W}$ to obtain the graph transition matrix:
\begin{equation}
\mathbf{A}
=
\mathbf{D}^{-1}\mathbf{W},
\end{equation}
where
$\mathbf{D}$
is the diagonal degree matrix with
$D_{ii}=\sum_j W_{ij}$. The resulting transition matrix $\mathbf{A}$ is used for Personalized PageRank relevance diffusion. The edge weights determine how relevance propagates across the document graph.
Higher edge weights indicate stronger structural or semantic associations and therefore allow more relevance flow during Personalized PageRank diffusion.

\subsection{Sparse-Dense Evidence Seeding}
\label{sec:seeding}

Given an online user query $Q$, the goal of this stage is to construct a query-dependent restart distribution vector $\mathbf{r} \in \mathbb{R}^{|\mathcal{V}|}$, which serves as the source signal for subsequent relevance diffusion. Rather than relying on a single granularity, EviProp initializes $\mathbf{r}$ by capturing both top-down global page layouts and bottom-up localized content cues.

\paragraph{Query Representation.}
We encode the textual query using the query encoder of the visual retriever and the text embedding model used for chunk 
retrieval:
\begin{equation}
    \mathbf{e}_Q^{\mathrm{txt}} = \operatorname{Enc}_{\mathrm{txt}}(Q), 
    \quad
    \mathbf{E}_Q^{\mathrm{vis}} = \operatorname{Enc}_{\mathrm{vis}}(Q),
\end{equation}
where $\mathbf{e}_Q^{\mathrm{txt}}$ is the query embedding from the text embedding model, and $\mathbf{E}_Q^{\mathrm{vis}}$ is the multi-vector query representation from the visual retriever.

\paragraph{Sparse Multimodal Chunk Seeds.}

To capture localized evidence that may be diluted during page-level matching, we compute a chunk relevance score for each content node $c \in \mathcal{C}$:
\begin{equation}
\label{eq:chunk_score}
s_{\mathrm{chunk}}(c) = 
\left\{
\begin{array}{l@{\hspace{-2.2em}}l}
\cos(\mathbf{e}_Q^{\mathrm{txt}}, \mathbf{t}_c), 
& c \in \mathcal{C}_{\mathrm{text}}, \\[6pt]
\alpha \cos(\mathbf{e}_Q^{\mathrm{txt}}, \mathbf{t}_c) 
& c \in \mathcal{C}_{\mathrm{vis}}. \\[3pt]
\quad + (1-\alpha)\operatorname{Sim}_{\mathrm{vis}}
(\mathbf{E}_Q^{\mathrm{vis}}, \mathbf{v}_c) &
\end{array}
\right.
\end{equation}

where $\mathbf{t}_c$ is the chunk text embedding, $\mathbf{v}_c$ denotes the visual token representations 
of the cropped visual region, and $\alpha \in [0,1]$ balances textual and visual relevance.
We then select the top-$K_c$ highest-scoring chunks as sparse evidence seeds:
\begin{equation}
    \mathcal{S}_c
    =
    \operatorname{TopK}
    \big(
    \{s_{\mathrm{chunk}}(c)\}_{c \in \mathcal{C}}
    \big).
\end{equation}

\paragraph{Dense Visual Page Priors.}

While chunk seeds provide precise localized evidence, page-level visual retrieval offers robust global coverage over the document.

For each page node $p \in \mathcal{P}$, we compute a dense visual relevance score:
\begin{equation}
    s_{\mathrm{vis}}(p)
    =
    \operatorname{Sim}_{\mathrm{vis}}
    (\mathbf{E}_Q^{\mathrm{vis}}, \mathbf{V}_p),
\end{equation}
where $\mathbf{V}_p$ denotes the multi-vector visual representation of page $p$. These scores serve as dense page-level priors in the restart distribution.

\paragraph{Restart Distribution Construction.}
Finally, we combine sparse chunk seeds and dense page priors into a unified restart distribution:
\begin{equation}
\label{eq:restart}
r(v)
\propto
\begin{cases}
s_{\mathrm{chunk}}(v), & v \in \mathcal{S}_c,\\
s_{\mathrm{vis}}(v),   & v \in \mathcal{P},\\
0,                     & \text{otherwise}.
\end{cases}
\end{equation}
The restart vector $\mathbf{r}$ is then $L_1$-normalized and serves as the initialization signal for subsequent Personalized PageRank relevance diffusion.

\subsection{Relevance Diffusion via Personalized PageRank}
\label{sec:ppr}
Given the restart distribution $\mathbf{r}$, EviProp performs relevance diffusion over the Chunk--Page graph using Personalized PageRank (PPR). Unlike independent page-level retrieval, this process enables relevance signals to propagate through document-internal structures, allowing localized chunk evidence, neighboring page context, and non-local semantic associations to jointly influence page relevance.

Let $\mathbf{A}$ denote the row-normalized transition matrix of the Chunk--Page graph. 
The propagated relevance vector 
$\boldsymbol{\pi}$ is iteratively updated as:
\begin{equation}
\label{eq:ppr}
\boldsymbol{\pi}^{(t+1)}
=
(1-\eta)\mathbf{r}
+
\eta \mathbf{A}^{\top}\boldsymbol{\pi}^{(t)},
\end{equation}
where $\eta \in (0,1)$ is the damping factor controlling the 
trade-off between restart preservation and graph propagation. 
Through this diffusion process, relevance flows from sparse 
chunk-level evidence to their parent pages, across sequentially 
adjacent pages, and between semantically or visually related 
regions. 

As a result, pages with weak initial visual similarity can still accumulate high relevance scores if they are strongly connected to evidence-bearing chunks or related pages. This allows EviProp to recover evidence pages that are under-ranked by independent page-wise matching.

\subsection{Final Evidence-Page Scoring}
\label{sec:scoring}
While graph diffusion improves evidence recovery, direct visual retrieval remains an important source of global query--page alignment. To preserve the robustness of the original visual retriever, we combine the propagated relevance score with the initial page-level visual similarity:
\begin{equation}
\label{eq:final_score}
\mathrm{Score}(p)
=
\gamma \cdot s_{\mathrm{vis}}(p)
+
(1-\gamma)\cdot \boldsymbol{\pi}_p,
\end{equation}
where $\gamma \in [0,1]$ balances direct visual matching and graph-propagated evidence relevance.

The final evidence-page set is obtained by ranking all pages according to $\mathrm{Score}(p)$:
\begin{equation}
\label{eq:topk}
\mathcal{P}_{r}
=
\operatorname{TopK}_{p \in \mathcal{P}}
\big(
\mathrm{Score}(p),\, K
\big),
\end{equation}
where $\mathcal{P}_{r}$ denotes the retrieved top-$K$ evidence pages.

Notably, EviProp operates purely at the retrieval stage and does not modify the downstream LVLM. The retrieved page images $\mathcal{P}_{r}$ are directly provided as visual context for answer generation:
\begin{equation}
\label{eq:lvlm}
\hat{y}
=
\operatorname{LVLM}
\left(
Q,\mathcal{P}_{r}
\right),
\end{equation}
where $\hat{y}$ denotes the final generated answer.

\section{Experiments}
\label{sec:experiments}

\subsection{Datasets and Evaluations}

\noindent\textbf{Datasets.}
We evaluate on \textbf{MMLongBench-Doc}~\cite{mmlongbench} and \textbf{LongDocURL}~\cite{longdocurl}, two benchmarks for long multimodal document understanding covering diverse topics and modalities. As shown in Table~\ref{tab:datasets}, the two datasets differ significantly in document length and information density.

\begin{table}[h]
\centering
\begin{small}
\resizebox{\columnwidth}{!}{%
    \begin{tabular}{lcccc}
    \toprule
    \textbf{Dataset} & \textbf{\# Question} & \textbf{\# Document} & \textbf{Avg.Pages} & \textbf{Avg.Tokens} \\
    \midrule
    MMLongBench-Doc & 1,082 & 135 & 47.5 & 24,992.6 \\
    LongDocURL  & 2,325 & 396 & 85.6 & 56,715.1 \\
    \bottomrule
    \end{tabular}%
}
\end{small}
\caption{Statistics of experimental datasets.}
\label{tab:datasets}
\end{table}

\noindent\textbf{Evaluation Metrics.}
Our primary evaluation focuses on evidence-page retrieval, where a method is judged by whether the retrieved page indices match annotated evidence pages.
We report Recall@K, Precision@K, NDCG@K, and MRR@K for $K \in \{1,3,5\}$ to measure evidence-page coverage and ranking quality.
For downstream QA, following the evaluation protocols of \cite{mmlongbench} and \cite{longdocurl}, we extract short answers from model outputs using GPT-4o~\cite{gpt} and report \textbf{Generalized Accuracy} with a rule-based evaluation script.

\subsection{Baselines}
\noindent\textbf{Retrieval Baselines.}
For evidence-page retrieval, we compare EviProp with the following representative methods:
\begin{itemize}[leftmargin=*]
\item \textbf{Text Chunk Retrieval}, which retrieves text chunks and maps them to their parent pages for page-level evaluation;
\item \textbf{Visual Retrieval}, which directly scores each page using ColPali-based query--page visual similarity;
\item \textbf{Text-Visual RRF}, which fuses text chunk ranking and visual page ranking using reciprocal rank fusion \cite{rrf}.
\end{itemize}

\noindent\textbf{Downstream QA Baselines.}
For downstream QA validation, we compare EviProp with the following representative methods:
\begin{itemize}[leftmargin=*]
\item \textbf{Text RAG}~\cite{rag-survery}, a standard OCR-based pipeline that retrieves textual chunks and feeds them to an LLM for answer generation;
\item \textbf{LVLM Direct Inference}~\cite{mmlongbench,longdocurl}, which directly feeds document page images into an LVLM without retrieval;
\item \textbf{M3DocRAG}~\cite{m3docrag}, which uses ColPali~\cite{colpali} as a page retriever to identify relevant pages and feeds only the retrieved page images to the LVLM for answer generation;
\item \textbf{MDocAgent}~\cite{mdocagent}, a collaborative multi-agent framework comprising specialized text and image agents for comprehensive document understanding.
\end{itemize}

\subsection{Implementation Details}
We use Qwen2.5-VL-7B~\cite{qwen2.5-vl} and Qwen3-VL-8B~\cite{qwen3vl} as LVLM backbones, and Qwen2.5-7B~\cite{qwen25} and Qwen3-8B~\cite{qwen3} as the corresponding LLMs for Text RAG baselines.
For plug-in evaluation, we integrate EviProp into MDocAgent~\cite{mdocagent} and MoLoRAG~\cite{molorag}, a logic-aware multimodal document retrieval framework, by replacing their original ColPali-based retrieval scores with EviProp final scores while keeping the remaining pipelines unchanged.
Detailed implementation settings for all baselines and EviProp are provided in Appendix~\ref{app:implementation}.

\section{Results and Analyses}

\subsection{Evidence-Page Retrieval Performance}
\label{sec:exp_retrieval}

\begin{table*}[t]
\centering
\small
\renewcommand{\arraystretch}{1.15}
\setlength{\tabcolsep}{7pt}
\definecolor{bgcolor}{rgb}{0.9, 0.95, 1.0}
\begin{tabular}{clcccccccc}
\toprule
\multirow{2}{*}{\textbf{Top-K}} & \multirow{2}{*}{\textbf{Method}} 
& \multicolumn{4}{c}{\textbf{MMLongBench-Doc}} 
& \multicolumn{4}{c}{\textbf{LongDocURL}} \\
\cmidrule(lr){3-6} \cmidrule(lr){7-10}
 & & \textbf{Recall} & \textbf{Precision} & \textbf{NDCG} & \textbf{MRR}
   & \textbf{Recall} & \textbf{Precision} & \textbf{NDCG} & \textbf{MRR} \\
\midrule

\multirow{4}{*}{\textbf{1}}
 & Text Chunk Retrieval  & 34.46 & 45.55 & 45.55 & 45.55 & 39.53 & 54.91 & 54.91 & 54.91 \\
 & Visual Retrieval      & 43.26 & 56.44 & 56.44 & 56.44 & 46.47 & 64.13 & 64.13 & 64.13 \\
 & Text-Visual RRF       & 39.34 & 51.41 & 51.41 & 51.41 & 44.03 & 60.81 & 60.81 & 60.81 \\
 \rowcolor{bgcolor}
 \cellcolor{white} & \textbf{EviProp (Ours)}
   & \textbf{45.41} & \textbf{59.67} & \textbf{59.67} & \textbf{59.67}
   & \textbf{47.43} & \textbf{65.55} & \textbf{65.55} & \textbf{65.55} \\
\midrule

\multirow{4}{*}{\textbf{3}}
 & Text Chunk Retrieval  & 53.31 & 26.00 & 50.37 & 53.59 & 61.10 & 30.79 & 57.95 & 64.20 \\
 & Visual Retrieval      & 64.15 & 31.58 & 61.48 & 65.01 & 66.87 & 33.66 & 64.95 & 72.23 \\
 & Text-Visual RRF       & 60.21 & 29.55 & 57.10 & 60.40 & 64.86 & 32.74 & 62.51 & 69.38 \\
 \rowcolor{bgcolor}
 \cellcolor{white} & \textbf{EviProp (Ours)}
   & \textbf{68.55} & \textbf{33.65} & \textbf{65.38} & \textbf{68.68}
   & \textbf{69.79} & \textbf{35.24} & \textbf{67.37} & \textbf{74.33} \\
\midrule

\multirow{4}{*}{\textbf{5}}
 & Text Chunk Retrieval  & 61.95 & 19.04 & 53.91 & 55.36 & 68.58 & 21.46 & 61.27 & 65.57 \\
 & Visual Retrieval      & 71.35 & 22.34 & 64.37 & 66.40 & 74.04 & 23.23 & 68.00 & 73.52 \\
 & Text-Visual RRF       & 68.77 & 21.15 & 60.48 & 61.85 & 72.23 & 22.76 & 65.81 & 70.70 \\
 \rowcolor{bgcolor}
 \cellcolor{white} & \textbf{EviProp (Ours)}
   & \textbf{75.68} & \textbf{23.47} & \textbf{68.08} & \textbf{69.83}
   & \textbf{76.36} & \textbf{24.06} & \textbf{70.21} & \textbf{75.52} \\
\bottomrule
\end{tabular}

\caption{Evidence-page retrieval performance comparison (in \%) under the top-$K$ setting. Best results are highlighted.}
\label{tab:retrieval}
\end{table*}

Table~\ref{tab:retrieval} evaluates whether each method retrieves the annotated evidence pages.
All methods involving visual retrieval use the same ColPali page-level scores for fair comparison.
The results lead to the following observations:
\paragraph{EviProp consistently improves evidence-page retrieval across datasets and retrieval budgets.}
Compared with the ColPali-based visual retrieval baseline, EviProp improves Top-3 Recall by $4.40$ points on MMLongBench-Doc and $2.92$ points on LongDocURL. This demonstrates that graph-based relevance diffusion helps recover evidence pages that are under-ranked by independent page-level matching.

\paragraph{EviProp also improves ranking quality beyond recall gains.}
On MMLongBench-Doc under Top-5 retrieval, EviProp improves NDCG by $3.71$ points and MRR by $3.43$ points over the visual retrieval baseline. This indicates that relevant evidence pages are not only retrieved more frequently, but also ranked earlier after propagation over the Chunk--Page graph.

\paragraph{Simple text--visual fusion is insufficient for evidence-page retrieval.}
Although Text-Visual RRF combines text and visual signals, it consistently underperforms pure visual retrieval on both benchmarks.
This suggests that simply adding text signals through rank fusion cannot effectively exploit document-internal associations and may introduce noisy local matches. In contrast, EviProp uses sparse chunk seeds as starting points for structured graph diffusion, allowing local evidence to propagate to related pages and chunks.

\begin{table*}[t]
\centering
\small
\renewcommand{\arraystretch}{1.12}
\setlength{\tabcolsep}{4pt}
\definecolor{bgcolor}{rgb}{0.93, 0.96, 1.0}
\newcommand{\hl}[1]{\cellcolor{bgcolor}#1}

\begin{tabular}{llcccccc}
\toprule
& & \multicolumn{3}{c}{\textbf{Qwen2.5-VL-7B Backbone}}
& \multicolumn{3}{c}{\textbf{Qwen3-VL-8B Backbone}} \\
\cmidrule(lr){3-5}
\cmidrule(lr){6-8}

\textbf{Category}
& \textbf{Method}
& \textbf{MMLong}
& \textbf{LongURL}
& \textbf{Avg.}
& \textbf{MMLong}
& \textbf{LongURL}
& \textbf{Avg.} \\

\midrule

\multirow{2}{*}{\textit{Standard Baselines}}
& Text RAG$^\dagger$
& 30.90 & 36.61 & 33.76
& 32.57 & 39.79 & 36.18 \\

& Direct Inference
& 32.77 & 26.38 & 29.58
& 44.88 & 30.24 & 37.56 \\

\midrule

\multirow{2}{*}{\textit{Multimodal RAG}}
& M3DocRAG
& 36.23 & 49.38 & 42.81
& 45.29 & 51.75 & 48.52 \\

& \hl{\textbf{EviProp (Ours)}}
& \hl{\textbf{38.34}}
& \hl{\textbf{50.65}}
& \hl{\textbf{44.50}}
& \hl{\textbf{47.75}}
& \hl{\textbf{53.52}}
& \hl{\textbf{50.64}} \\

\midrule

\multirow{4}{*}{\textit{Plug-in Integration}}
& MDocAgent (Original ColPali)
& 38.00 & 46.91 & 42.46
& 46.80 & 52.91 & 49.86 \\

& \hl{MDocAgent (w/ EviProp Retriever)}
& \hl{\textbf{39.01}}
& \hl{\textbf{50.61}}
& \hl{\textbf{44.81}}
& \hl{\textbf{49.09}}
& \hl{\textbf{53.65}}
& \hl{\textbf{51.37}} \\

\cmidrule(lr){2-8}

& MoLoRAG (Original ColPali)
& 38.17 & 50.49 & 44.33
& 47.38 & 53.21 & 50.30 \\

& \hl{MoLoRAG (w/ EviProp Retriever)}
& \hl{\textbf{39.27}}
& \hl{\textbf{51.17}}
& \hl{\textbf{45.22}}
& \hl{\textbf{48.69}}
& \hl{\textbf{55.52}}
& \hl{\textbf{52.11}} \\

\bottomrule
\end{tabular}

\caption{
\textbf{Comprehensive downstream QA accuracy comparison (\%) under the Top-3 retrieval setting.}
We evaluate EviProp as both a standalone multimodal retriever and a plug-in retrieval module integrated into existing multimodal QA frameworks.
$^\dagger$ Text RAG uses Qwen2.5-7B and Qwen3-8B as text-only LLM backbones.
}
\label{tab:master_qa_results}
\end{table*}

\subsection{Downstream QA Validation}
\label{sec:exp_qa}

We further evaluate whether improved evidence-page retrieval transfers to downstream QA. Since EviProp directly optimizes the retrieval stage rather than the LVLM generation process, 
QA accuracy is reported as downstream validation. As shown in Table~\ref{tab:master_qa_results}, we draw the following conclusions:

\paragraph{EviProp consistently outperforms independent visual retrieval as a standalone retriever.}
Despite only modifying the retrieval stage, EviProp surpasses M3DocRAG by up to $2.12$ points on average and achieves comparable or higher accuracy than MDocAgent, a multi-agent framework with substantially higher inference cost.

\paragraph{Retrieval improvements transfer to downstream answer accuracy.}
These results suggest that improving evidence-page retrieval can lead to consistent downstream QA gains, although final answer correctness remains affected by the LVLM’s visual perception and reasoning ability.

\begin{table}[t]
\centering
\begin{small}
\resizebox{\columnwidth}{!}{%
\begin{tabular}{lcccc}
\toprule
\multirow{2}{*}{\textbf{Variant}} 
& \multicolumn{2}{c}{\textbf{MMLongBench-Doc}} 
& \multicolumn{2}{c}{\textbf{LongDocURL}} \\
\cmidrule(lr){2-3} \cmidrule(lr){4-5}
& \textbf{R@3} & \textbf{NDCG@3} & \textbf{R@3} & \textbf{NDCG@3} \\
\midrule
\rowcolor{bgcolor}
\textbf{EviProp (Full)} & \textbf{68.55} & \textbf{65.38} & \textbf{69.79} & \textbf{67.37} \\
\midrule
w/o PPR Diffusion        & 64.15 & 61.48 & 66.87 & 64.95 \\
Pure PPR Score           & 67.10 & 63.65 & 68.37 & 65.21 \\
\midrule
w/o Vis\&Seq Edges & 67.30 & 63.99 & 68.48 & 66.94 \\
\midrule
w/o Page Seeds           & 64.24 & 60.56 & 67.82 & 64.02 \\
w/o Chunk Seeds          & 64.30 & 61.85 & 66.52 & 65.04 \\
\bottomrule
\end{tabular}
}
\end{small}
\caption{
Ablation studies on EviProp components.
We report Recall@3 and NDCG@3 for evidence-page retrieval.
}
\label{tab:ablation}
\end{table}

\subsection{Retrieval Mechanism Analysis}
\label{sec:analysis}

To validate the effectiveness of each component in EviProp, we conduct ablation studies on both MMLongBench-Doc and LongDocURL.
Since downstream QA uses the top-3 retrieved pages, we report Recall@3 and NDCG@3 under the same retrieval budget.
Results are summarized in Table~\ref{tab:ablation}.

\paragraph{Effect of relevance diffusion.}
Removing PPR diffusion leads to a consistent drop in both Recall@3 and NDCG@3, showing that independent page matching cannot fully recover evidence pages supported by local or structural cues.
Pure PPR Score ranks pages using only the propagated posterior $\pi_p$, without interpolation with the original visual prior.
It maintains relatively high recall but lowers NDCG@3, suggesting that diffusion alone may over-propagate relevance without sufficient anchoring to direct visual matching.
The final scoring function therefore combines propagated relevance with the original visual prior.

\paragraph{Effect of document-internal edges.}
Removing visual similarity and sequential page--page edges causes a moderate but consistent drop across both benchmarks.
This indicates that page-level relational edges provide complementary propagation paths beyond chunk--page membership.

\paragraph{Effect of sparse-dense seeding.}
Both page seeds and chunk seeds contribute to retrieval performance.
Removing either type of seed from the PPR restart distribution leads to clear drops in Recall@3 and NDCG@3.
This shows that dense page-level visual priors help keep diffusion anchored to global query--page alignment, while sparse chunk seeds inject fine-grained local evidence into the graph.

\begin{table}[t]
\centering
\small
\renewcommand{\arraystretch}{1.12}
\setlength{\tabcolsep}{4.2pt}
\definecolor{bgcolor}{rgb}{0.93,0.96,1.0}

\begin{tabular}{lcccc}
\toprule

\multirow{2}{*}{\textbf{Retriever}}
& \multicolumn{2}{c}{\textbf{MMLongBench-Doc}}
& \multicolumn{2}{c}{\textbf{LongDocURL}} \\

\cmidrule(lr){2-3}
\cmidrule(lr){4-5}

& \textbf{R@3} & \textbf{R@5}
& \textbf{R@3} & \textbf{R@5} \\

\midrule

ColPali
& 68.18 & 74.74
& 69.48 & 76.64 \\

\rowcolor{bgcolor}
EviProp
& \textbf{70.17}
& \textbf{77.41}
& \textbf{71.79}
& \textbf{77.98} \\

\bottomrule
\end{tabular}

\caption{\textbf{Plug-in retriever evaluation within MoLoRAG.}
Replacing the original ColPali retrieval scores with EviProp consistently improves recall.}

\label{tab:molorag_integration_clean}
\end{table}

\subsection{Integration with Existing Retrieval Pipelines}
\label{sec:integration}

We further evaluate whether EviProp can serve as a plug-in retrieval module for existing document QA pipelines.
We integrate EviProp into MoLoRAG and MDocAgent by replacing their ColPali retrieval scores with EviProp final scores, 
while keeping the remaining pipelines unchanged.
As shown in Table~\ref{tab:molorag_integration_clean}, integrating EviProp into MoLoRAG consistently improves 
evidence-page recall under both Recall@3 and Recall@5 on both benchmarks.
These retrieval gains further translate into downstream QA improvements (Table~\ref{tab:master_qa_results}), suggesting 
that EviProp can serve as a compatible plug-in retrieval module for existing document QA pipelines.

\subsection{Efficiency Analysis}
\label{sec:efficiency}

\begin{figure}[t]
  \centering
  \includegraphics[width=\linewidth]{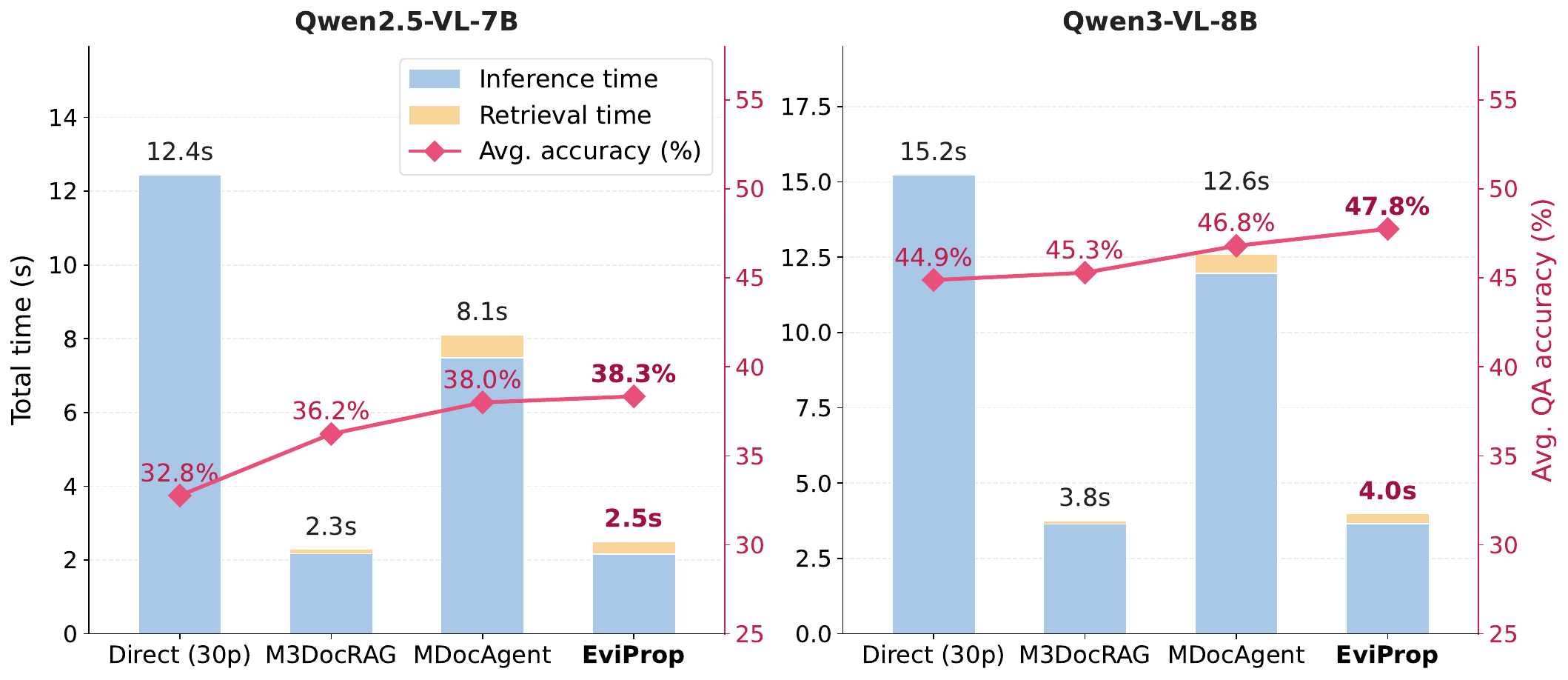}
  \caption{Latency and QA accuracy comparison on MMLongBench-Doc. EviProp achieves the highest accuracy with latency comparable to M3DocRAG.}
  \label{fig:efficiency}
\end{figure}

Figure~\ref{fig:efficiency} compares end-to-end latency and downstream QA accuracy on MMLongBench-Doc. EviProp achieves the best accuracy under both backbones while maintaining latency comparable to M3DocRAG. Compared with M3DocRAG, EviProp adds only 0.2s of retrieval overhead from graph diffusion, accounting for a small fraction of total latency. MDocAgent incurs substantially higher latency due to multi-turn agent interactions, yet achieves lower accuracy than EviProp. Direct inference over 30 pages is both slow and accuracy-limited due to context truncation. Thus, EviProp achieves a favorable accuracy--efficiency trade-off by shifting reasoning from slow autoregressive generation to fast retrieval-time propagation.

\section{Conclusion}
\label{sec:conclusion}

We present EviProp, a retrieval method for evidence-page discovery in visually rich long documents. 
EviProp constructs a multimodal Chunk--Page graph and performs seeded relevance diffusion from dense visual page priors and sparse chunk seeds, enabling localized evidence and document-internal associations to jointly influence page relevance. 
Experiments on MMLongBench-Doc and LongDocURL show that EviProp consistently improves both evidence-page retrieval 
and downstream QA accuracy, while adding negligible online retrieval overhead.

\section*{Limitations}
\label{sec:limitations}
While EviProp demonstrates consistent gains across retrieval budgets and evaluation benchmarks, future work may explore 
query-adaptive edge weighting and more selective seeding strategies to further improve robustness in documents with 
repetitive or template-like elements.
In addition, downstream QA accuracy is also influenced by the LVLM's visual perception and reasoning ability, leaving joint 
optimization of retrieval and generation as a promising direction for future exploration.

\bibliography{main}


\FloatBarrier
\newpage

\clearpage
\appendix


\section{Use of Large Language Models}
The research presented in this paper, including the core ideas, experimental design, and quantitative results, is the original work of the authors. A large language model was used as a writing assistant for tasks such as polishing prose, improving clarity, and correcting grammatical errors in the manuscript. All final content was reviewed and edited by the authors to ensure it accurately reflects our research and contributions.

\section{Algorithm Pseudocode}
\label{app:pseudocode}

Algorithm~\ref{alg:eviprop} summarizes the complete EviProp pipeline.
The procedure consists of two phases: an offline graph construction phase and an 
online retrieval phase. In the offline phase, EviProp parses the document into page 
nodes and content nodes, and builds a multimodal Chunk--Page graph encoding 
hierarchical, sequential, and similarity relations. In the online phase, given a query, 
EviProp initializes a restart distribution from sparse chunk seeds and dense page 
priors, runs Personalized PageRank to diffuse relevance over the graph, and returns 
the top-$K$ pages by combining the diffused posterior with the visual prior. The retrieved pages are then passed to a downstream LVLM as visual context for answer generation.

\begin{algorithm}[ht]
\small
\caption{EviProp: Seeded Relevance Diffusion for Evidence-Page Retrieval}
\label{alg:eviprop}
\begin{algorithmic}[1]
\REQUIRE Document $D=\{p_i\}_{i=1}^{M}$, query $Q$, retrieval budget $K$
\ENSURE Generated answer $\hat{y}$
\STATE $(\mathcal{P}, \mathcal{C}, \rho) \leftarrow \mathrm{Parse}(D)$ 
\COMMENT{$\rho(c)$ denotes the parent page of chunk $c$}
\STATE $\mathcal{V} \leftarrow \mathcal{P} \cup \mathcal{C}$
\STATE \textcolor{gray}{\textit{// Offline Chunk--Page graph construction}}
\STATE $\mathcal{E}_{\mathrm{hier}} \leftarrow \{(c,\rho(c)),(\rho(c),c) \mid c \in \mathcal{C}\}$
\STATE $\mathcal{E}_{\mathrm{seq}} \leftarrow \{(p_i,p_{i+1}),(p_{i+1},p_i)\}_{i=1}^{M-1}$
\STATE $\mathcal{E}_{\mathrm{sim}} \leftarrow \mathrm{VisSim}(\mathcal{P}) \cup \mathrm{SemSim}(\mathcal{C})$
\STATE $\mathcal{E} \leftarrow \mathcal{E}_{\mathrm{hier}} \cup \mathcal{E}_{\mathrm{seq}} \cup \mathcal{E}_{\mathrm{sim}}$
\STATE $\mathbf{W} \leftarrow \mathrm{Weight}(\mathcal{V},\mathcal{E})$
\STATE $\mathbf{A} \leftarrow \mathbf{D}^{-1}\mathbf{W}$
\STATE \textcolor{gray}{\textit{// Online sparse-dense evidence seeding}}
\FORALL{$c \in \mathcal{C}$}
    \STATE $s_{\mathrm{chunk}}(c) \leftarrow \mathrm{ChunkScore}(Q,c)$
\ENDFOR
\STATE $\mathcal{S}_c \leftarrow \mathrm{TopK}_{K_c}(\mathcal{C}, s_{\mathrm{chunk}})$
\FORALL{$p \in \mathcal{P}$}
    \STATE $s_{\mathrm{vis}}(p) \leftarrow \mathrm{Norm}(\mathrm{PageScore}(Q,p))$
\ENDFOR
\FORALL{$v \in \mathcal{V}$}
    \STATE $r(v) \leftarrow
    \begin{cases}
    \max(0, s_{\mathrm{chunk}}(v)), & v \in \mathcal{S}_c,\\
    s_{\mathrm{vis}}(v), & v \in \mathcal{P},\\
    0, & \text{otherwise}.
    \end{cases}$
\ENDFOR
\STATE $\mathbf{r} \leftarrow \mathbf{r}/\|\mathbf{r}\|_1$
\STATE \textcolor{gray}{\textit{// Seeded relevance diffusion}}
\STATE $\boldsymbol{\pi} \leftarrow \mathrm{PPR}(\mathbf{A}, \mathbf{r}, \eta)$
\COMMENT{$\boldsymbol{\pi}=(1-\eta)\mathbf{r}+\eta\mathbf{A}^{\top}\boldsymbol{\pi}$}
\STATE \textcolor{gray}{\textit{// Final evidence-page scoring}}
\FORALL{$p \in \mathcal{P}$}
    \STATE $\mathrm{Score}(p) \leftarrow \gamma s_{\mathrm{vis}}(p) + (1-\gamma)\boldsymbol{\pi}_p$
\ENDFOR
\STATE $\mathcal{P}_r \leftarrow \mathrm{TopK}_{K}(\mathcal{P}, \mathrm{Score})$
\STATE \textcolor{gray}{\textit{// Downstream answer generation}}
\STATE $\hat{y} \leftarrow \mathrm{LVLM}(Q, \mathcal{P}_r)$
\RETURN $\hat{y}$
\end{algorithmic}
\end{algorithm}

\section{Experimental and Implementation Details}
\label{app:implementation}

All experiments were conducted on a cluster equipped with \textbf{4 NVIDIA RTX 4090 (24GB)} GPUs and \textbf{4 NVIDIA A100 (80GB)} GPUs.
For fair comparison, all methods use the same page images rasterized from PDFs at \textbf{144\,DPI} with PyMuPDF\footnote{\url{https://pymupdf.readthedocs.io}}; structured text is extracted with MinerU~\cite{mineru}.

\subsection{Details of Retrieval Metrics}
\label{app:metrics}

We use standard retrieval metrics to evaluate evidence-page retrieval, including Recall@K, Precision@K, NDCG@K, and MRR@K.
For each query, let the set of annotated evidence pages be $\mathcal{P}_{gt}=\{p^{1}_{gt},\ldots,p^{n}_{gt}\}$, where $n=|\mathcal{P}_{gt}|$.
Let the ranked top-$K$ retrieved pages be $\mathcal{P}_{r}^{K}=\{p^{1}_{r},\ldots,p^{K}_{r}\}$.
We use an indicator function $\mathbb{I}(\cdot)$ that returns $1$ if the condition holds and $0$ otherwise.
All metrics are first computed per query and then averaged over the evaluation set.

\paragraph{Recall@K.}
Recall@K measures the proportion of gold evidence pages covered by the top-$K$ retrieved pages:
\begin{equation}
    \mathrm{Recall@K}
    =
    \frac{1}{n}
    \sum_{i=1}^{K}
    \mathbb{I}\!\left(p^{i}_{r}\in\mathcal{P}_{gt}\right).
\end{equation}

\paragraph{Precision@K.}
Precision@K measures the proportion of retrieved pages that are gold evidence pages:
\begin{equation}
    \mathrm{Precision@K}
    =
    \frac{1}{K}
    \sum_{i=1}^{K}
    \mathbb{I}\!\left(p^{i}_{r}\in\mathcal{P}_{gt}\right).
\end{equation}

\paragraph{NDCG@K.}
NDCG@K measures ranking quality under binary relevance.
A retrieved page is assigned relevance $1$ if it belongs to $\mathcal{P}_{gt}$ and $0$ otherwise.
The discounted cumulative gain is:
\begin{equation}
    \mathrm{DCG@K}
    =
    \sum_{i=1}^{K}
    \frac{
    \mathbb{I}\!\left(p^{i}_{r}\in\mathcal{P}_{gt}\right)
    }{
    \log_{2}(i+1)
    }.
\end{equation}
The ideal discounted cumulative gain is:
\begin{equation}
    \mathrm{IDCG@K}
    =
    \sum_{i=1}^{\min(n,K)}
    \frac{1}{\log_{2}(i+1)}.
\end{equation}
Then NDCG@K is defined as:
\begin{equation}
    \mathrm{NDCG@K}
    =
    \frac{\mathrm{DCG@K}}{\mathrm{IDCG@K}}.
\end{equation}

\paragraph{MRR@K.}
MRR@K measures the reciprocal rank of the first retrieved evidence page within the top-$K$ results:
\begin{equation}
    \mathrm{MRR@K} = \frac{1}{\min\{i \mid p^{i}_{r}\in\mathcal{P}_{gt},\, i\le K\}},
\end{equation}
where $\mathrm{MRR@K}=0$ if no evidence page appears in the top-$K$ results.

\paragraph{Handling queries without annotated evidence pages.}
For retrieval-stage evaluation, we exclude queries without annotated evidence pages.
For downstream QA evaluation, we follow the official benchmark protocol and evaluate on the full test set.

\subsection{Baseline Implementation Settings}
\label{app:baseline_settings}

\begin{itemize}[leftmargin=*]
    \item \textbf{Text RAG:}
    We use MinerU~\cite{mineru} to extract text from the original documents and segment it into chunks of \textbf{1,200 characters} with a \textbf{200-character} overlap.
    We embed each text chunk using OpenAI \texttt{text-embedding-3-large} and perform dense retrieval to construct the textual context for the downstream LLM.

    \item \textbf{LVLM Direct Inference:}
    Due to the context limits of the backbone LVLMs, we truncate the input to the \textbf{first 30 pages} and feed the page images directly to Qwen2.5-VL and Qwen3-VL backbones without retrieval.

    \item \textbf{M3DocRAG:}
    We follow the method and official repository of M3DocRAG.
    For retrieval, we use \textbf{ColPali}~\cite{colpali} as the visual document encoder.
    We evaluate M3DocRAG with the same LVLM backbones used in our experiments, namely Qwen2.5-VL-7B and Qwen3-VL-8B, to ensure comparability.

    \item \textbf{MDocAgent:}
    We follow the official implementation of MDocAgent, using \textbf{Colpali}~\cite{colpali} for image retrieval and \textbf{ColBERTv2}~\cite{colbertv2} for text retrieval.
    The original \textbf{LLaMA-3.1-8B}~\cite{llama} is kept as the LLM for the text agent.
    We evaluate the pipeline with Qwen2.5-VL-7B and Qwen3-VL-8B as the LVLM backbone for the remaining agents.
\end{itemize}

\subsection{EviProp Implementation Settings}
\label{app:eviprop_settings}

\begin{itemize}[leftmargin=*]
    \item \textbf{Parsing and encoders:}
    Page images are rasterized using PyMuPDF at 144\,DPI. Structured text and visual regions are extracted using MinerU~\cite{mineru}. Text is segmented into chunks of \textbf{1,200 characters} with a \textbf{200-character} overlap and embedded using OpenAI \texttt{text-embedding-3-large}.
    
    For visual pages, we use ColPali~\cite{colpali} embeddings.
    MinerU is also used to automatically crop local visual regions, such as figures and tables, as visual chunks.
    We generate captions for these visual chunks using GPT-4o and treat the captions as their textual representations.

    \item \textbf{Graph construction:}
    We construct one Chunk--Page graph for each document.
    The graph contains chunk--page membership edges, sequential page edges, page--page visual similarity edges, and chunk--chunk semantic similarity edges.
    We set the chunk semantic similarity threshold $\tau_{\mathrm{txt}}=0.5$, the hierarchical edge weight $w_{\mathrm{inc}}=5.0$, and the sequential page edge weight $w_{\mathrm{seq}} = 0.5$.
    Detailed edge weighting schemes are provided in Appendix~\ref{app:graph_weights}.

    \item \textbf{Visual Similarity Computation.}
    $\operatorname{Sim}_{\mathrm{vis}}$ is implemented as ColPali-style late 
    interaction~\cite{colbert,colpali} followed by robust min--max normalization. 
    Concretely, for a query $Q$ and a visual unit $X$ (either a page or a visual chunk):
    \begin{equation}
        \operatorname{Sim}_{\mathrm{vis}}(Q, X)
        =
        \frac{s^{\mathrm{raw}} - s_{\min}}
             {\max(s_{\max} - s_{\min},\; R)},
    \end{equation}
    where
    \begin{equation}
        s^{\mathrm{raw}} = \sum_{i=1}^{|E_Q|}\max_{j\in[|E_X|]}\left(\mathbf{q}_i^\top \mathbf{x}_j\right)
    \end{equation}
    is the raw MaxSim score, $s_{\min}$ and $s_{\max}$ are the minimum and maximum 
    raw scores across all pages in the document, and $R=10.0$ is a normalization 
    floor that prevents over-compression when the score range is small.

    \item \textbf{Seeding and scoring:}
    We use top-$K_c=3$ chunk seeds.
    For visual chunks, the textual--visual fusion weight is set to $\alpha=0.7$, meaning 70\% weight on the textual score and 30\% on the visual score.
    The final page scoring weight is set to $\gamma=0.5$.
    We use a PPR damping factor of $\eta=0.5$ and stop diffusion when
    $\lVert\boldsymbol{\pi}^{(t+1)}-\boldsymbol{\pi}^{(t)}\rVert_1<10^{-6}$.
\end{itemize}

\subsection{EviProp Graph Edge Weighting Details}
\label{app:graph_weights}

This section provides detailed edge weighting schemes for 
the Chunk--Page graph.

\paragraph{Page--page visual similarity edges.}
For each page $p_i$, we compress its multi-vector visual 
representation $\mathbf{V}_{p_i}$ into a single unit vector 
by mean pooling followed by $\ell_2$ normalization:
\begin{equation}
    \mathbf{u}_{p_i}
    =
    \frac{\mathrm{mean}(\mathbf{V}_{p_i})}
         {\|\mathrm{mean}(\mathbf{V}_{p_i})\|_2}.
\end{equation}
The visual edge weight is defined as a clipped cosine similarity:
\begin{equation}
    w_{\mathrm{vis}}(p_i,p_j)
    =
    \max\!\bigl(0,\; \mathbf{u}_{p_i}^{\top}\mathbf{u}_{p_j}\bigr).
\end{equation}

\paragraph{Chunk--chunk semantic similarity edges.}
Given two chunk nodes $c_i$ and $c_j$, we retain semantic 
edges whose cosine similarity exceeds a threshold 
$\tau_{\mathrm{txt}}$. The edge weight is defined as:
\begin{equation}
    w_{\mathrm{sim}}(c_i,c_j)
    =
    [\max(0,\cos(\mathbf{t}_{c_i}, \mathbf{t}_{c_j}))]^3,
\end{equation}
where cubic scaling reduces the influence of weak semantic 
overlaps and emphasizes stronger chunk-level associations.

\paragraph{Edge weight merging.}
If multiple edge types connect the same pair of nodes 
$(v_i, v_j)$, we retain the largest weight:
\begin{equation}
    W_{ij} \leftarrow \max\bigl(W_{ij},\, w(v_i, v_j)\bigr).
\end{equation}

\subsection{Plug-in Integration and Efficiency Measurement}
\label{app:integration}

\paragraph{MoLoRAG integration.}
For integration with MoLoRAG~\cite{molorag}, we replace its ColPali-based page similarity scores with EviProp scores during retrieval, while keeping the page graph, VLM relevance scoring, and beam-search fusion pipeline unchanged.

\paragraph{MDocAgent integration.}
For integration with MDocAgent~\cite{mdocagent}, we replace its visual retriever scores with EviProp final scores to 
retrieve evidence pages, while keeping the remaining agentic pipeline unchanged.

\paragraph{Efficiency measurement.}
Latency is measured as the average time per query on MMLongBench-Doc.
We separately report retrieval time, downstream inference time, and total time.

\begin{figure*}[t]
  \centering
  \includegraphics[width=0.95\textwidth]{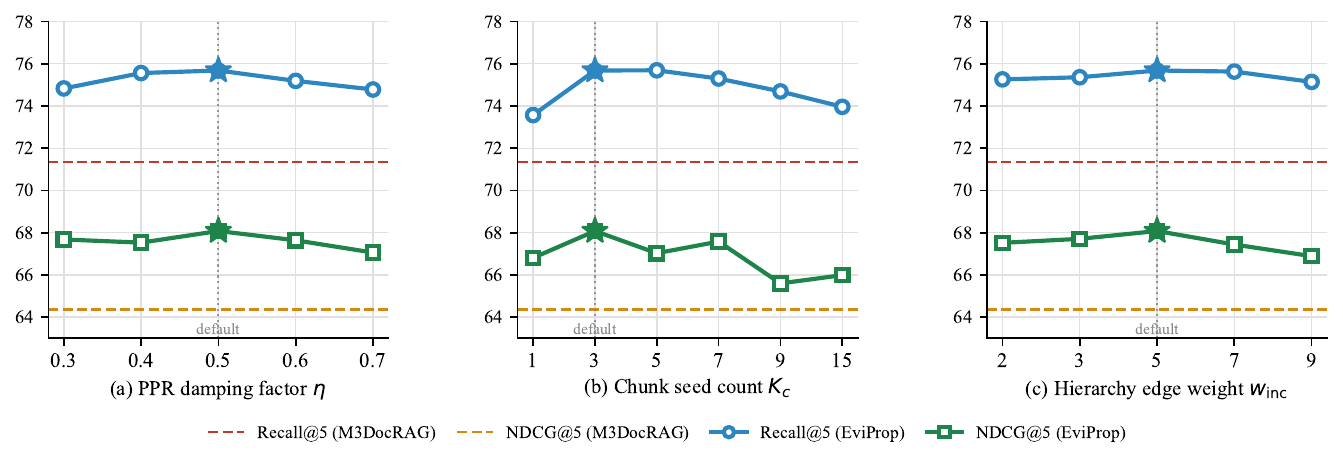}
  \caption{
  Hyperparameter sensitivity analysis on MMLongBench-Doc.
  We report Recall@5 and NDCG@5 under different values of the PPR damping factor $\eta$, chunk seed count $K_c$, and hierarchy edge weight $w_{\mathrm{inc}}$.
  }
  \label{fig:sensitivity}
\end{figure*}

\section{Additional Experimental Results}
\label{app:additional_experiments}

\subsection{Hyperparameter Sensitivity}
\label{app:sensitivity}

We analyze the sensitivity of EviProp to key hyperparameters, including the PPR damping factor $\eta$, the number of chunk seeds $K_c$, and the hierarchy edge weight $w_{\mathrm{inc}}$.
As shown in Figure~\ref{fig:sensitivity}, EviProp remains consistently above the Visual Retrieval baseline across a range of hyperparameter settings.
This indicates that the retrieval gains do not rely on a single carefully tuned configuration.

\begin{figure*}[t!]
  \centering
  \includegraphics[width=\textwidth]{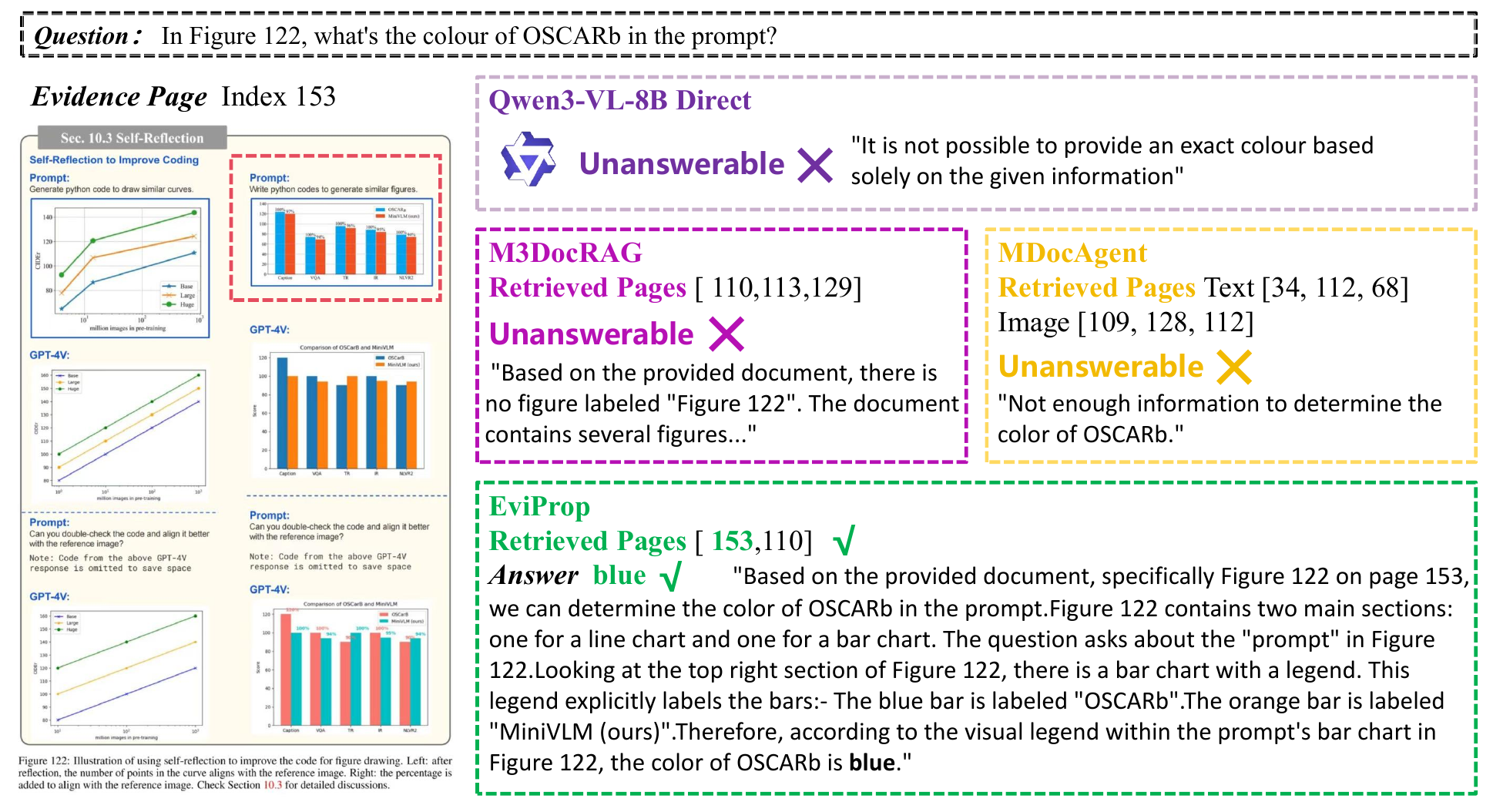} 
  \caption{
  \textbf{Case study on MMLongBench-Doc.}
  EviProp retrieves the correct evidence page by leveraging seeded relevance diffusion, enabling the LVLM to provide the correct answer.
  In contrast, LVLM Direct Inference and baseline retrieval methods fail due to limited input context or irrelevant retrieved pages.
  }
  \label{fig:case_study}
\end{figure*}

\subsection{Additional Results on UDA}
\label{app:uda}

We further evaluate EviProp on PaperTab and FetaTab, two table-centric subsets of the UDA benchmark~\cite{hui2024uda}, derived from NLP research papers and Wikipedia tables respectively, both targeting question answering over table-rich documents.
Following the evaluation protocol of MDocAgent, we use GPT-4o as the evaluator and assign a binary correctness score for each answer.
As shown in Table~\ref{tab:uda}, EviProp consistently outperforms both M3DocRAG and MDocAgent on both subsets under both backbones.

\begin{table}[t]
\centering
\definecolor{bgcolor}{rgb}{0.9, 0.95, 1.0}
\resizebox{\columnwidth}{!}{%
\begin{tabular}{llcc}
\toprule
\textbf{Model} & \textbf{Method} & \textbf{PaperTab} & \textbf{FetaTab} \\
\midrule
\multirow{3}{*}{Qwen2.5-VL-7B}
 & M3DocRAG & 28.50 & 63.58 \\
 & MDocAgent & 29.52 & 66.04 \\
 \rowcolor{bgcolor}
 \cellcolor{white} & \textbf{EviProp (Ours)} & \textbf{30.79} & \textbf{66.54} \\
\midrule
\multirow{3}{*}{Qwen3-VL-8B}
 & M3DocRAG & 38.68 & 63.78 \\
 & MDocAgent & 39.69 & 66.14 \\
 \rowcolor{bgcolor}
 \cellcolor{white} & \textbf{EviProp (Ours)} & \textbf{40.71} & \textbf{66.63} \\
\bottomrule
\end{tabular}%
}
\caption{QA accuracy (\%) on UDA subsets (PaperTab and FetaTab) under the Top-3 retrieval setting.}
\label{tab:uda}
\end{table}

\section{Case Study}
\label{app:case_study}

Figure~\ref{fig:case_study} presents a case study on MMLongBench-Doc requiring the identification of a specific visual attribute in a targeted chart.
LVLM Direct Inference marks the question as ``unanswerable'' due to limited input context.
Baseline retrieval methods such as M3DocRAG and MDocAgent fail to locate the evidence page, leading to incorrect answers.
In contrast, EviProp retrieves the correct evidence page by leveraging seeded relevance diffusion over document-internal associations.
This precise grounding enables the LVLM to read the visual content and answer the question correctly.

\section{Prompts for Visual Content Description}
\label{app:prompts}

To bridge the modality gap between textual queries and visual document elements (e.g., charts, figures, and complex tables), we use a Large Vision-Language Model (LVLM) to generate detailed textual descriptions for visual chunks ($\mathcal{C}_{\text{vis}}$).
These descriptions serve as the textual representation $\mathbf{t}_c$ for visual nodes, enabling the computation of semantic similarity scores $s_{\text{chunk}}(c)$ during sparse seeding.
We use separate prompts for figures/charts and tables.

\begin{tcolorbox}[title={Figure/Chart Description Prompt}]
\textbf{System Instruction:} Please analyze this figure/chart in detail and provide a comprehensive description.

\textbf{Task Requirements:} Provide a comprehensive and detailed visual description following these guidelines:
\begin{itemize}[leftmargin=1.2em, nosep]
    \item Describe the overall composition and layout.
    \item Identify all objects, text, and visual elements.
    \item Explain relationships between elements.
    \item Note colors, visual style, and design patterns.
    \item Describe any actions or data flows shown.
    \item Include technical details (charts, diagrams, architectures, etc.).
    \item For charts/graphs: describe axes, data trends, and comparisons.
    \item Always use specific names instead of pronouns.
\end{itemize}

\textbf{Input Context:}
\begin{itemize}[leftmargin=1.2em, nosep]
    \item Path: \texttt{\{image\_path\}}
    \item Caption: \texttt{\{caption\}}
    \item Footnotes: \texttt{\{footnotes\}}
\end{itemize}

\textbf{Output Constraint:} Please provide a detailed description (200--500 words).
\end{tcolorbox}

\begin{tcolorbox}[title={Table Description Prompt}]
\textbf{System Instruction:} Please analyze this table content and provide a comprehensive description.

\textbf{Task Requirements:} Provide a comprehensive analysis of the table including:
\begin{itemize}[leftmargin=1.2em, nosep]
    \item Table structure and organization.
    \item Column headers and their meanings.
    \item Row labels and categories.
    \item Key data points and patterns.
    \item Statistical insights and trends.
    \item Relationships between data elements.
    \item Significance of the data presented.
    \item Always use specific names and values instead of general references.
\end{itemize}

\textbf{Input Context:}
\begin{itemize}[leftmargin=1.2em, nosep]
    \item Path: \texttt{\{image\_path\}}
    \item Caption: \texttt{\{caption\}}
    \item Body text visible: \texttt{\{table\_body\}}
    \item Footnotes: \texttt{\{footnotes\}}
\end{itemize}

\textbf{Output Constraint:} Please provide a detailed description (200--500 words).
\end{tcolorbox}

\end{document}